\documentclass[12pt]{article}
\usepackage{graphicx}
\usepackage{cite}
\newcommand{\DP}{\Delta\Pi}
\newcommand{\ind}[2]{^{\mbox{\scriptsize $#1$}}_{\mbox{\scriptsize #2}}}
\newcommand{\inds}[2]{^{\mbox{\scriptsize $#1$}}_{\mbox{\tiny #2}}}
\newcommand{\nf}{n_{\mbox{\scriptsize f}}}
\newcommand\AMH[1]{\Delta\alpha\ind{#1}{had}}

\begin{document}

\begin{center}

{\Large\bf Hadronic contributions to electroweak observables in the framework of~DPT

}

\vskip5mm

{\large A.V.~Nesterenko}

{\small\it BLTPh, Joint Institute for Nuclear Research,
Dubna, 141980, Russian Federation}

\end{center}

\vskip2.5mm

\noindent
\centerline{\bf Abstract}

\vskip2.5mm

\centerline{\parbox[t]{125mm}{%
The~hadronic contributions to the muon anomalous magnetic moment and to
the shift of the electromagnetic fine structure constant at the scale of
$Z$~boson mass are evaluated within dispersively improved perturbation
theory~(DPT). The~latter merges the corresponding perturbative input with
intrinsically nonperturbative constraints, which originate in the
respective kinematic restrictions. The~obtained results conform with
recent assessments of the quantities on hand.}}

\vskip5mm

Certain nonperturbative information on the low--energy hadron dynamics is
embodied within dispersion relations. In~particular, the latter render the
kinematic restrictions on the relevant physical processes into the
mathematical form and impose intrinsically nonperturbative constraints on
such quantities as the hadronic vacuum polarization function~$\Pi(q^2)$,
$R$--ratio of electron--positron annihilation into hadrons~$R(s)$, and the
Adler function~$D(Q^2)$. These constraints have been merged with
corresponding perturbative input within dispersive approach to
QCD~\cite{DPT1a, PRD88, JPG42QCD15} (its preliminary formulation was
discussed in Refs.~\cite{DPTPrelim1, DPTPrelim2}), which provides the
unified integral representations for the functions on hand:
\begin{eqnarray}
\label{PDPT}
\DP(q^2,\, q_0^2) \!\!\!\!& = &\!\!\!\! \DP^{(0)}(q^2,\, q_0^2) +
\int_{m^2}^{\infty} \rho(\sigma)
\ln\biggl(\frac{\sigma-q^2}{\sigma-q_0^2}
\frac{m^2-q_0^2}{m^2-q^2}\biggr)\frac{d\,\sigma}{\sigma}, \\
\label{RDPT}
R(s) \!\!\!\!& = &\!\!\!\! R^{(0)}(s) + \theta(s-m^2) \int_{s}^{\infty}\!
\rho(\sigma) \frac{d\,\sigma}{\sigma}, \\
\label{DDPT}
D(Q^2) \!\!\!\!& = &\!\!\!\! D^{(0)}(Q^2) +\frac{Q^2}{Q^2+m^2}
\int_{m^2}^{\infty} \rho(\sigma)
\frac{\sigma-m^2}{\sigma+Q^2} \frac{d\,\sigma}{\sigma}.
\end{eqnarray}
Here $\DP(q^2\!,\, q_0^2) = \Pi(q^2) - \Pi(q_0^2)$, $m^2 = 4m_{\pi}^2$,
$\theta(x)$~is the unit step--function [$\theta(x)=1$ if $x \ge 0$ and
$\theta(x)=0$ otherwise], the leading--order terms read~\cite{Feynman,
QEDAB}
\begin{eqnarray}
\label{P0L}
\DP^{(0)}(q^2,\, q_0^2) \!\!\!\!& = &\!\!\!\!
2\,\frac{\varphi - \tan\varphi}{\tan^3\varphi}
- 2\,\frac{\varphi_{0} - \tan\varphi_{0}}{\tan^3\varphi_{0}}, \\
\label{R0L}
R^{(0)}(s) \!\!\!\!& = &\!\!\!\! \theta(s - m^2)\bigl[1-(m^2/s)\bigr]^{3/2}, \\
\label{D0L}
D^{(0)}(Q^2) \!\!\!\!& = &\!\!\!\! 1 + 3\bigl[1 - \sqrt{1 + \xi^{-1}}\,
\sinh^{-1}\bigl(\xi^{1/2}\bigr)\bigr]\xi^{-1},
\end{eqnarray}
$\sin^2\!\varphi = q^2/m^2$, $\sin^2\!\varphi_{0} = q^{2}_{0}/m^2$,
$\xi=Q^2/m^2$, see Refs.~\cite{DPT1a, PRD88, JPG42QCD15} for the details.
The perturbative part of the spectral density~$\rho(\sigma)$ entering
Eqs.~(\ref{PDPT})--(\ref{DDPT}) can be expressed in terms of the strong
correction to the Adler function
\begin{equation}
\label{RhoPert}
\rho\ind{}{pert}(\sigma) = \lim_{\varepsilon \to 0_{+}}
\bigl[d\ind{}{pert}(-\sigma - i \varepsilon) -
d\ind{}{pert}(-\sigma + i \varepsilon)\bigr]/(2 \pi i),
\end{equation}
as well as in terms of the strong corrections to the functions~$\Pi(q^2)$
and~$R(s)$. The integral representations (\ref{PDPT})--(\ref{DDPT})
contain no unphysical singularities and substantially extend the range of
applicability of QCD perturbation theory towards the infrared domain.

In~particular, the dispersive approach to QCD enables one to describe OPAL
(update~2012, Ref.~\cite{OPAL9912}) and ALEPH (update~2014,
Ref.~\cite{ALEPH0514}) experimental data on inclusive $\tau$~lepton
hadronic decay in vector and axial--vector channels in a self--consistent
way~\cite{PRD88, QCD14} (see also Refs.~\cite{DPT3, C12}). Additionally,
the representations~(\ref{PDPT})--(\ref{DDPT}) conform with the results of
Bethe--Salpeter calculations~\cite{PRL99PRD77} as well as of lattice
simulations~\cite{RCTaylor}. The Adler function~(\ref{DDPT}) agrees with
its experimental prediction in the entire energy range~\cite{DPT1a, DPT1b,
DPT2}.

The integral representations (\ref{PDPT})--(\ref{DDPT}) along with the
respective perturbative input~(\ref{RhoPert}) constitute the
``dispersively improved perturbation theory''~(DPT) expressions for the
functions on hand. At the one--loop level the spectral
function~(\ref{RhoPert}) assumes a simple form, specifically,
\mbox{$\rho\ind{(1)}{pert}(\sigma) = (4/\beta_{0})
[\ln^{2}(\sigma/\Lambda^2)+\pi^2]^{-1}$} ($\beta_{0} = 11 - 2\nf/3$,
$\nf$~is the number of active flavors, and~$\Lambda$ is the QCD scale
parameter), whereas at the higher loop levels Eq.~(\ref{RhoPert}) becomes
rather cumbersome, see Refs.~\cite{Review, CPC, APTnum} for the details.

\begin{figure}[t]
\centerline{\includegraphics[width=75mm]{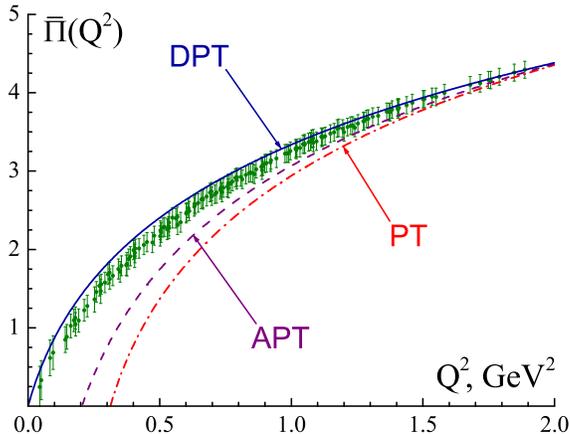}}
\caption{Hadronic vacuum polarization function in the framework of
various approaches: DPT~expression (solid curve), APT~prediction
(dashed curve), perturbative approximation (dot--dashed curve),
and lattice data~\cite{Lat5}~(circles).}
\label{Plot:PDPT}
\end{figure}

It is worth noting that in the massless limit $(m=0)$ for the case of
perturbative spectral function~(\ref{RhoPert}) Eqs.~(\ref{RDPT})
and~(\ref{DDPT}) become identical to those of the ``analytic perturbation
theory''~(APT)~\cite{APT} (see also Refs.~\cite{APT1, APT2, APT3, APT4,
APT5a, APT5b, APT6, APT7, APT8, APT9, APT11}). However, the massless limit
ignores some of the nonperturbative constraints, which relevant dispersion
relations impose on the functions on hand, that appears to be substantial
for the studies of hadron dynamics at low energies, see Refs.~\cite{PRD88,
DPT1a, JPG42QCD15, C12, DPT2}.

For practical purposes it is convenient to deal with the subtracted at
zero form of Eq.~(\ref{PDPT}), namely
\begin{equation}
\label{PDPT2}
\bar{\Pi}(Q^2) = \DP(0,-Q^2) = \DP^{(0)}(0, -Q^2) +
\int_{m^2}^{\infty} \rho(\sigma)
\ln\biggl(\frac{1+Q^2/m^2}{1+Q^2/\sigma}\biggr)
\frac{d\,\sigma}{\sigma}.
\end{equation}
As one can infer from Fig.~\ref{Plot:PDPT}, the DPT expression for the
hadronic vacuum polarization function~(\ref{PDPT2}) contains no unphysical
singularities and proves to be in a good agreement with lattice
data~\cite{Lat5} (the rescaling procedure described in
Refs.~\cite{RhoRescale1, RhoRescale2} was applied). The presented result
corresponds to the four--loop level, $\Lambda=419\,$MeV, and~$\nf=2$.
Figure~\ref{Plot:PDPT} also displays the one--loop Eq.~(\ref{PDPT}) in the
massless limit (which corresponds to APT) as well as the one--loop
perturbative approximation of~$\Pi(q^2)$. However, the latter is
inapplicable at low energies due to unphysical singularities, whereas the
APT prediction for~$\Pi(q^2)$ diverges at~$q^2 \to 0$, that invalidates it
in the infrared domain, too.

\begin{figure}[t]
\centerline{\includegraphics[width=75mm,clip]{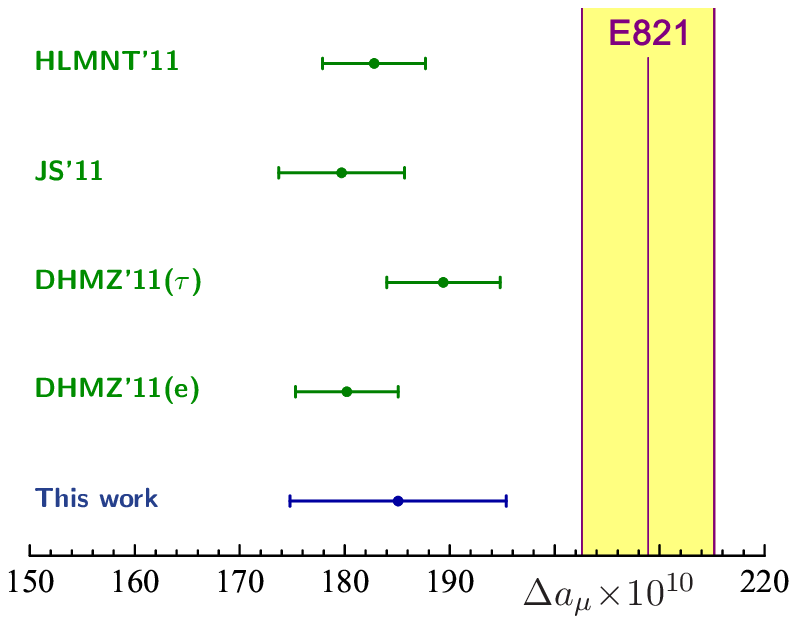}
\hfill
\raisebox{-0.6pt}{\includegraphics[width=75mm,clip]{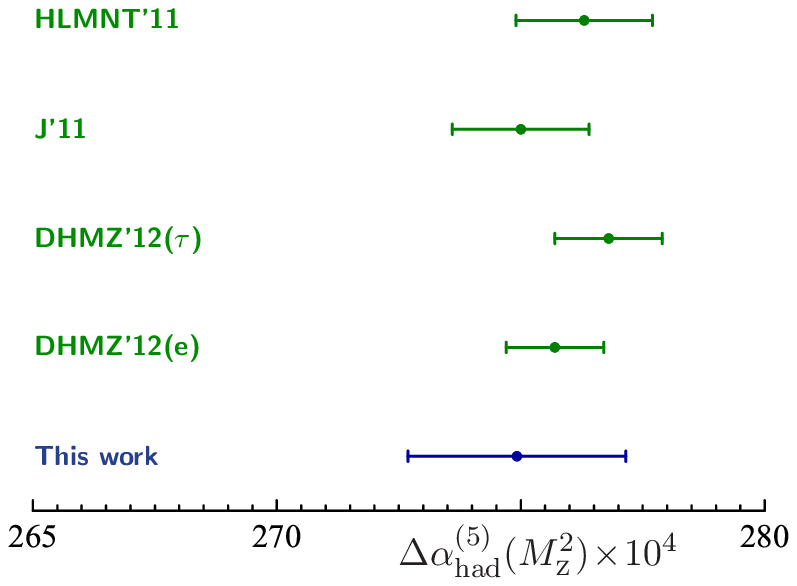}}}
\caption{Left--hand plot: theoretical evaluations (circles) and
experimental measurement (shaded band) of the subtracted muon anomalous
magnetic moment ($\Delta a_{\mu} = a_{\mu} - a_{0}$, $\,a_{0} = 11659
\times 10^{-7}$). Right--hand plot: theoretical evaluations of hadronic
contribution to the shift of electromagnetic fine structure constant at
the scale of $Z$~boson mass.}
\label{Plot:EW}
\end{figure}

The DPT expression for the hadronic vacuum polarization
function~(\ref{PDPT2}), being applicable in the entire energy range,
enables one to perform the assessment of the hadronic contributions to
electroweak observables without involving experimental data
on~\mbox{$R$--ratio}. In~particular, the four--loop DPT prediction of the
value of the leading--order hadronic contribution to the muon anomalous
magnetic moment~\cite{JPG42QCD15}
\begin{equation}
\label{AmuHVP}
a_{\mu}^{\mbox{\tiny HLO}} =
\frac{1}{3} \biggl(\frac{\alpha}{\pi}\biggr)^{\!2}
\!\int_{0}^{1}\!(1-x)
\bar{\Pi}\biggl(\!m_{\mu}^{2}\,\frac{x^2}{1-x}\!\biggr) dx =
(696.1 \pm 9.5) \times 10^{-10},
\end{equation}
appears to be in a good agreement with its recent
estimations~\cite{HLMNT11, JS11, DHMZ11}. The complete muon anomalous
magnetic moment~$a_{\mu}$ includes the QED contribution~\cite{AmuQED}, the
electroweak contribution~\cite{AmuEW}, as well as the
higher--order~\cite{HLMNT11} and light--by--light~\cite{AmuHlbl} hadronic
contributions, that, together with $a_{\mu}^{\mbox{\tiny
HLO}}$~(\ref{AmuHVP}) yields $a_{\mu} = (11659185.1 \pm 10.3) \times
10^{-10}$. The obtained~$a_{\mu}$ corresponds to two standard deviations
from the experimental measurement $a_{\mu}^{\mbox{\scriptsize exp}} =
(11659208.9 \pm 6.3) \times 10^{-10}$~\cite{MuonExp2} and, as one can
infer from Fig.~\ref{Plot:EW}, conforms with its recent
evaluations~\cite{HLMNT11, JS11, DHMZ11}, see Ref.~\cite{JPG42QCD15} for
the details.

Another quantity of an apparent interest is the hadronic contribution to
the electromagnetic running coupling
\begin{equation}
\label{AMH}
\AMH{}(q^2) = - \frac{\alpha}{3\pi}\,q^2
\,\mathcal{P}\!\!\int_{m^2}^{\infty}\!
\frac{R(s)}{s-q^2}\frac{d\,s}{s}.
\end{equation}
The four--loop DPT prediction for the five--flavor hadronic contribution
to the shift of the electromagnetic fine structure constant at the scale
of $Z$~boson mass~\cite{JPG42QCD15}
\begin{equation}
\label{AMH_DPT}
\AMH{(5)}(M\inds{2}{Z}) = (274.9 \pm 2.2) \times 10^{-4}
\end{equation}
agrees with its recent evaluations~\cite{HLMNT11, DHMZ11, J11},
see~Fig.~\ref{Plot:EW}. In~turn, Eq.~(\ref{AMH_DPT}) together with
leptonic~\cite{AEMLep} and top~quark~\cite{AEMtop} contributions leads to
$\alpha\ind{-1}{em}(M\inds{2}{Z}) = 128.962 \pm 0.030$, that also conforms
with recent assessments of the quantity on hand~\cite{HLMNT11, DHMZ11,
J11}, see Ref.~\cite{JPG42QCD15} for the details.

\end{document}